\pdfoutput=1
\documentclass[aps,american,pra,citeautoscript,floatfix,pra,pdftex,superscriptaddress,%
longbibliography,notitlepage,%
]{revtex4-1}
\usepackage{amsfonts,amsmath,amssymb}
\usepackage[T1]{fontenc}
\usepackage[utf8]{inputenc}
\usepackage{graphicx}
\usepackage{microtype}
\usepackage[obeyFinal,textsize=footnotesize]{todonotes}
\usepackage{stackrel}
\usepackage{graphicx}
\usepackage{hyperref}
\usepackage{enumerate,tensor,xfrac}
\usepackage{nicefrac}
\usepackage{bm}
\usepackage{hyperref}
\usepackage{color}
\hypersetup{%
  pdfpagemode=None, 
  pdfstartpage=1,
  pdfstartview=FitH,
  pdfmenubar=true,
  pdftoolbar=true,
  colorlinks = true,
  linkcolor=blue,
  citecolor=blue,
  bookmarksopen=false
}
\usepackage{soul,xcolor}
\usepackage[normalem]{ulem}

\usepackage{float}

\usepackage{amsmath}
\usepackage{graphicx}
\newcommand{\eg}{e.\,g.}%
\newcommand{\ie}{i.\,e.}%
\newcommand{\ry}{Rydberg }

\newcommand{\bea}{\begin{eqnarray}}
\newcommand{\ea}{\end{eqnarray}}
\newcommand{\eea}{\end{eqnarray}}

\newcommand{\sumint}[1]

\begin{document}
\title{A protocol to realize triatomic ultralong range \ry molecules in an ultracold KRb gas}

\author{Rosario Gonz\'alez-F\'erez}

\affiliation{Instituto Carlos I de F\'isica Te\'orica y Computacional and Departamento de F\'isica At\'omica, Molecular y Nuclear, Universidad de Granada, 18071 Granada, Spain}
\author{Seth T. Rittenhouse}

\affiliation{Department of Physics, The United States Naval Academy, Annapolis, Maryland 21402, USA}
\author{Peter Schmelcher}

\affiliation{Zentrum f\"ur Optische Quantentechnologien, Universit\"at Hamburg, Luruper Chaussee 149, 22761 Hamburg, Germany}
\affiliation{The Hamburg Centre for Ultrafast Imaging, Universit\"at Hamburg, Luruper Chaussee 149, 22761 Hamburg, Germany}
\author{H. R. Sadeghpour}

\affiliation{ITAMP, Harvard-Smithsonian Center for Astrophysics 60 Garden St., Cambridge, Massachusetts 02138, USA}


\vspace{10pt}

\begin{abstract}
We propose an experimentally realizable scheme to produce triatomic ultralong-range \ry molecules (TURM), formed in ultracold KRb traps. A near resonant coupling of the non-zero quantum defect Rydberg levels with the KRb molecule in N=0 and N=2 rotational levels, is engineered which exploits the unique Rydberg electron-molecule anisotropic dipole interaction. This near resonant coupling enhances the TURM binding and produces favorable Franck-Condon factors. Schemes for both postassium and rubidium excitations are demonstrated. 
\end{abstract}

\maketitle

%
%
%
%

\section{Introduction}
\label{sec:intro}

Rydberg atoms are superb probes of their environments. Spectral shift and broadening of Rydberg lines in thermal gases have been used for temperature and density diagnostics of atomic and molecular laboratory and astrophysical gases \cite{Allard1982,Szudy1996,amaldi1934}, and were instrumental in the early development of low-energy atomic collision physics \cite{fermi1934}. The advent of ultracold trapping and cooling techniques now permit routine probes of environmental bath in the quantum limit \cite{Lukin2001,Schmidt2016,Camargo2018,Saffmann2010}.

Ultracold molecules offer another intriguing toolkit for manipulation and control of interatomic and atom-molecule interactions~\cite{cold_mol}
Ultracold molecules possessing permanent electric dipole moments (PEDM) are particularly in vogue for a host of applications in ultracold chemistry, quantum information processing and many-body quantum physics.

The interaction of a Rydberg atom with a non-degenerate ground state atom is best described as the quasi-free electron scattering at low energy from pertuber atoms. When the electron de Broglie wavelength is large compared with other interaction length scales, the Fermi pseudopotential method was invoked for binding of a Rydberg atom with a bath atom, forming Rydberg molecules bound at ultra-long separations \cite{greene00,hamilton02,chibisov02,bendkowsky09}. When the environmental gas is comprised of molecules possessing  permanent electric field dipole moments, the low-energy Rydberg electron scattering is best described as the charge-dipole interaction \cite{fermi47}. This interaction has a critical value- the Fermi-Teller critical dipole moment $d_c\approx 1.63$~Debye; molecules with supercritical dipole moments may bind electrons and create anions and those with subcritical dipole moments scatter electrons~\cite{Crawford1967,Clark1979,Sadeghpour2000}. 

The original proposal for creation of ultralong range triatomic Rydberg molecules (TURM) comprised of a Rydberg atom and a polar molecule \cite{rittenhouse10}, imagined a two-level polar molecule, say the opposite-parity electronic doublet in OH interacting with a Rydberg electron. Subsequent calculations expanded upon the original proposal to successively incorporate rotational molecules and their linear response with electric fields \cite{rittenhouse11,gonzalez15,mayle12}. Interesting aspects of these molecules are  their prodigious permanent electric dipole moments, sensitivity to small external electric field, and the existence of multiple, state-dependent orientations of the  molecules~\cite{mayle12}.  The full rotational level structure in the ground state polar molecule was incorporated in~\cite{gonzalez15,aguilera15}.

Ultralong range triatomic molecules discussed here and above have utility beyond Rydberg chemistry, such as coherently oriented molecules, and for quantum information processing and simulation of magnetic impurity interactions. Specific examples include Rydberg mediated conditional molecular gates and their non-destructive readouts~\cite{Kuznetsova2011,Kuznetsova2016} and simulation of indirect spin-spin interactions~\cite{Kuznetsova2018} in bilayers of Rydberg atoms and rotational polar molecules. In many such examples, the existence of large dipole moments is a blessing~\cite{Ni2018}.

Here, we detail with accurate calculations, an unexplored near resonant scheme between rotational and Rydberg levels,  which we show is most optimal for experimental realization.  In all previous calculations, the most interesting triatomic molecular states were those associated with the nearly degenerate manifold of high angular momentum Rydberg states.  In each case, the molecules were predicted to have little to no $s$- or $d$-wave Rydberg character \cite{mayle12,gonzalez15} making them inaccessible to conventional two-photon Rydberg excitation schemes.  We leverage the rotational structure in the ground vibrational state of KRb molecules to find advantageous $s$ or $d$ Rydberg orbital admixtures for the experimental realization of ultralong range triatomic Rydberg molecules.

We consider a mixture of ultracold potassium $^{39}$K and rubidium atoms
$^{87}$Rb, where ultracold 
extremely weakly bound Feshbach  
$^{39}$K$^{87}$Rb molecules are formed by magnetoassociation~\cite{ni08}.
These extremely weakly bound molecules are transferred to the electronic and vibrational ground-state by a coherent two-photon Raman scheme, which creates KRb molecules in either 
$N=0$ or  $N=2$ rotational state~\cite{ni08}. 
The rotational constant of KRb is 
$B= 1.114$~GHz~\cite{ni09}, and its electric dipole moment is subcritical, $d=0.566$~Debye.

In such a  atom-molecule mixture, either potassium or rubidium atoms could be excited into $nd$ or $ns$  \ry states by standard two-photon 
absorption schemes. If the KRb molecule is found in the Rydberg orbit, the TURM molecules, K-KRb or Rb-KRb, were predicted to 
form~\cite{rittenhouse10,rittenhouse11,mayle12}. In the previous attempts,~\cite{gonzalez15,aguilera15}, 
we showed 
that the adiabatic electronic potential curves (APC) evolving from the $nd$ or $ns$ \ry thresholds are deep enough 
to accommodate several vibrational bound states, but the corresponding vibrational spacing to be smaller
than $1$~MHz, and therefore beyond spectroscopic resolution. In addition, the potential energy surfaces associated with these states are likely to be highly sensitive to the $s$-wave scattering length associated with the electron-KRb polarization interaction, a quantity that is currently not known. 

In contrast, the  adiabatic potential energy curves evolving from the \ry 
degenerate manifolds, \ie, $l\gtrsim 3$, exhibit potential wells with few GHz depths~\cite{gonzalez15}, 
and, the corresponding vibrational bound states can be resolved spectroscopically. Furthermore, the electron-molecule interaction is dominated by the charge-dipole interaction. The \ry degenerate 
manifold could be accessed experimentally from the neighbouring $ns$ or $nd$ \ry state, allowing for the creation of the TURM molecules. 

\section{The theoretical protocol}
\label{sec:protocol}
Our aim is to suggest an experimentally viable scheme for the production of the TURM in an adiabatic electronic potential evolving from a $ns$ or $nd$ \ry state, 
by coupling these electronic states to those evolving from a hydrogenic-like degenerate manifold.
\begin{figure*}[t]
\centering
 \includegraphics[scale=1]{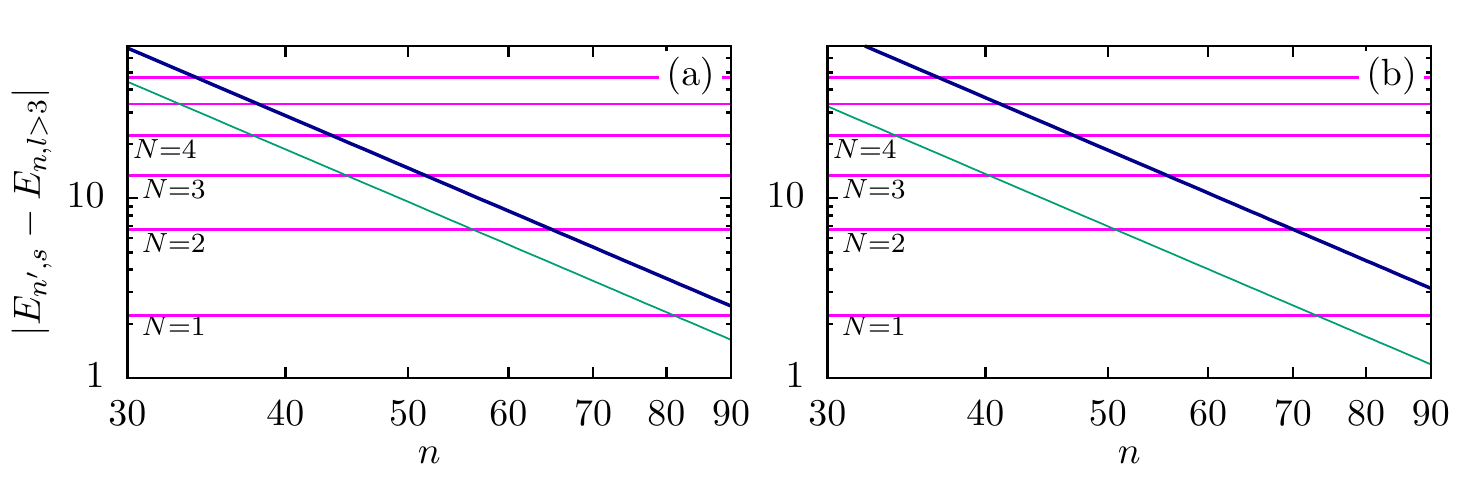}
\caption{(a) For potassium, 
energy splittings in GHz of the \ry states K$((n+2)s)$ (thin green line)   and K$(nd)$ (thick blue line) 
from the neighbouring  \ry manifold K$(n,l\ge3)$ as a function of the principal quantum number $n$. 
(b)  For Rubidium
energy splittings in GHz of  the \ry states Rb$((n+3)s)$ (thin green line) and Rb$((n+1)d)$ (thick blue line) from the neighbouring 
\ry manifold Rb$(n,l\ge3)$ versus  the principal quantum number $n$.
The horizontal lines represent the energies of the rotationally excited states of KRb.
\label{fig:K_splitting}}
\end{figure*}
The goal is to enhance the depth of these APC, and, therefore, the energy spacing of the vibrational 
bound states, by coupling these quantum defect states to the closest degenerate manifold using the 
dipole interaction with the polar diatomic molecule. 

To engineer  such a coupling, the energy splitting of the $ns$ or $nd$ \ry state from the neighbouring degenerate manifold, \ie, $|E(n's)-E(n,l>3)|$ or $|E(n'd)-E(n,l>3)|$, should be close to  
but larger than the energy of a given rotational excited state of the field-free KRb molecule. Thus, and taking as an 
example the excitation of potassium, the APC of the K$(n+2,s)$-KRb($N>0$) or K$(n,d)$-KRb($N>0$) 
TURM should be immersed among the APCs  of  K$(n,l>3)$-KRb($N=0$), \ie, the APC 
evolving from the \ry  
degenerate manifold with principal quantum number $n$ and KRb in its rotational ground-state $N=0$. 
We note that due to interaction with the core electrons, Rydberg valence electron energies are shifted from the hydrogenic levels; the usual terminology is that the levels obtain a quantum defect $E_{nl}=-\frac{R_\infty}{(n-\mu_{l})}$, where $R_\infty$ is the Rydberg constant and $\mu_l$ is the $l$-dependent quantum defect. With each additional orbital angular momentum, $\mu_l$ becomes smaller such that for $l$ larger than some value, here $l\geq 3$, $\mu_l\sim 0$; hence the degenerate manifold.

For potassium,~\autoref{fig:K_splitting}~(a) shows the energy shifts $|E(n+2,s)-E(n,l>3)|$ and 
$|E(n,d)-E(n,l>3)|$ versus the quantum number $n$. In this figure, the rotational excited states of KRb are represented by 
the horizontal lines. The possible \ry manifold candidates to engineer the coupling are given by the 
principal quantum number $n$ having an energy slight larger than one of the horizontal lines. For instance, 
the $51\lesssim n \lesssim 56$ ($59\lesssim n \lesssim 64$) degenerate manifolds of potassium are good candidates to create the TURM via the K$((n+2)s)$ (K$(nd)$) quantum-defect  \ry state and 
KRb($N=2$). 
The corresponding results for exciting the rubidium atom are shown in~\autoref{fig:K_splitting}~(b). 
For Rb, the ($45\lesssim n \lesssim 50$) ($64\lesssim n \lesssim 69$) degenerate manifolds are good  
candidates to create the TURM via the dipole coupling of the Rb$((n+3)s)$ (Rb$((n+1)d)$) \ry state and 
KRb($N=2$). 

This paper is organized as follows. In~\autoref{sec:hamiltonian} we present the theoretical method.
The APC of two \ry molecules K-KRb and  Rb-KRb are discussed and analyzed in~\autoref{sec:results}.  
Our conclusions are provided in~\autoref{sec:con}.

\section{The TURM Hamiltonian}
\label{sec:hamiltonian}
We consider a  triatomic molecule (TURM) formed by  a Rydberg atom and a ground state heteronuclear 
diatomic molecule. The diatomic molecule is described  within the  Born-Oppenheimer and 
rigid rotor  approximations. The \ry atom is represented by an effective single-electron system.
In this framework, the adiabatic Hamiltonian is 
\begin{equation}
\label{eq:Hamil_adiabatic}
H_{ad}=H_A+H_{mol}, 
\end{equation}
where $H_A$  represents  the single electron Hamiltonian describing the \ry atom 
\begin{equation}
\label{eq:Hamil_atom}
H_A=-\frac{\hbar^2}{2m_e}\nabla^2_{r}+V_l(r), 
\end{equation}
where $V_l(r)$ is the  $l$-dependent model potential~\cite{marinescu94}, with $l$ being the angular momentum
 quantum number of the \ry electron with respect to its positively charged core.

The molecular Hamiltonian describing the  rigid rotor molecule  and  the charge-dipole interaction reads 
\begin{equation}
\label{eq:Hamil_molecule}
H_{mol}=B\mathbf{N}^2
-\mathbf{d}\cdot\mathbf{F}_{ryd}(\mathbf{R},\mathbf{r})
\end{equation}
with $B$ being the rotational constant, $\mathbf{N}$ the molecular angular momentum operator and 
$\mathbf{d}$ the permanent electric dipole moment of the diatomic molecule. 
$\mathbf{F}_{ryd}(\mathbf{R},\mathbf{r})$ is the electric field due to the Rydberg
electron and the ion at  position $\mathbf{R}$  
\begin{equation}
\label{eq:field_rydberg_e_core}
\mathbf{F}_{ryd}(\mathbf{R},\mathbf{r})=e\frac{\mathbf{R}}{R^3}+e\frac{\mathbf{r}-\mathbf{R}}{|\mathbf{r}-\mathbf{R}|^3}
\end{equation}
where $e$ is the electron charge, $\mathbf{r}$ is the position of the \ry electron, and 
$\mathbf{R}$ is the  position of the diatomic molecule with respect to the \ry core. 
The KRb molecule is in its ground electronic state, i. e. $^1\Sigma_g$, whose electronic spin is zero and therefore does not introduce spin-exchange mixing of the electronic states.  The presence of both hyperfine and electron-quadrupole interactions will introduce only a weak mixing and small shifts in the APCs and are therefore neglected here. The 
rovibrational motion is treated in the rigid rotor approximation and the rotational constant is $B=1.114$ GHz.

We consider the  position of the \ry core  fixed at the center of the coordinate 
system of the laboratory fixed frame, and the
diatomic molecule is located at the $Z$-axis, \ie, $\mathbf{R}=R\hat{Z}$.
By fixing the distance between the diatomic molecule and the \ry core $R$, we solve the Schr\"odinger 
equation associated with the Hamiltonian~\eqref{eq:Hamil_adiabatic} and obtain the adiabatic electronic 
potential  curves (APC) as a function of $R$. In the limit of very large separations, the interaction 
between the \ry atom and KRb tends to zero, and the APC approaches the energies of the two
non-interacting species, \ie,  $E_{n,l}+N(N+1)B$, with $E_{n,l}$ being the energy of the corresponding \ry 
state, and $N(N+1)B$ the rotational energy of the KRb molecule in its electronic and vibrational ground-state. 

To solve the Schr\"odinger equation associated with the Hamiltonian~\eqref{eq:Hamil_adiabatic},
we perform a basis set expansion 
in terms of the coupled basis
\begin{equation}
\Psi(\mathbf{r},\Omega;R)=\sum_{n,l,N,J} C_{n,l,m}^J(R)\Psi_{nl,N}^{JM_J}(\mathbf{r},\Omega)
\label{eq:basis_expansion}
\end{equation}
where  the sum in  $J$ satisfies  $|l-N|\le J\le l+N$, and 
\begin{equation}
 \Psi_{nl,N}^{JM_J}(\mathbf{r},\Omega)
 =\sum_{m_l=-l}^{m_l=l}\sum_{M_N=-N}^{M_N=N}
\langle l m_l NM_N| J M_J\rangle
Y_{NM_N}(\Omega)\,
\psi_{nlm}(\mathbf{r}) 
\label{eq:coupled_basis}
\end{equation}
with $\langle l m_l N M_N| J M_J\rangle$ being the Clebsch-Gordan coefficient,
$J=|l-N|,\dots,l+N$,  and  $M_J=-J,\dots,J$. 
$\psi_{nlm}(\mathbf{r})$ is the \ry electron wave function
with $n$, $l$ and $m$ being the principal, orbital and magnetic quantum numbers, respectively,
and $Y_{NM_{N}}(\Omega)$ is the field-free rotational wave function of the diatomic molecule.
The total angular momentum of the \ry molecule,  excluding an overall rotation, is given by $\mathbf{J}=\mathbf{l}+\mathbf{N}$,
where $\mathbf{l}$ is the orbital angular momentum of the \ry electron,
and $\mathbf{N}$ is the molecular rotational 
angular momentum of the diatomic molecule.
Here, we include rotational excitations of KRb up to $N\le 5$, and for the \ry wave functions,
we include the degenerate manifold with $l>3$, and the neighbouring  quantum defect states with 
$l=0,1,2,3$. 
In this work, we focus on the states with $M_J=0$, \ie, $M_N+m_l=0$,
and the  basis is formed by $2142$ and $2632$
coupled states  for $n=46$ and $n=56$, respectively.

To gain physical insight into the features of these exotic molecules, we compute their electric dipole 
moment as 
\begin{equation}
D_{ryd}(R)=
\langle\Psi|r\cos\theta_e|\Psi\rangle=\int
\Psi^*(\mathbf{r},\Omega;R)
r\cos \theta_e
\Psi(\mathbf{r},\Omega;R)d\mathbf{r} d\Omega
\label{eq:expec_dipole_moment}
\end{equation}
where $\theta_e$ is the polar angle of the \ry electron. This integral  is non-zero if there are 
partial waves with different parity, \ie,  $\Delta l=\pm 1$, 
contributing to the wave function~\eqref{eq:basis_expansion}.
We also explore the orientation and alignment of the KRb molecule within the TURM given by
\begin{equation}
O_{ryd}(R)=
\langle\Psi|\cos\theta|\Psi\rangle=\int
\Psi^*(\mathbf{r},\Omega;R)
\cos \theta
\Psi(\mathbf{r},\Omega;R)d\mathbf{r} d\Omega
\label{eq:expec_coseno}
\end{equation}
\begin{equation}
A_{ryd}(R)=
\langle\Psi|\cos^2\theta|\Psi\rangle=\int
\Psi^*(\mathbf{r},\Omega;R)
\cos^2 \theta
\Psi(\mathbf{r},\Omega;R)d\mathbf{r} d\Omega
\label{eq:expec_coseno2}
\end{equation}
The weight  of a certain rotational state of KRb in the TURM wave function~\eqref{eq:basis_expansion} 
reads
\begin{equation}
{\cal C}_N(R)=\sum_{n,l,J}|C_{n,l,m}^J(R)|^2 \qquad \text{satisfying} \qquad
\sum_{N}{\cal C}_N(R)=1, 
\label{eq:weight_N}
\end{equation} 
and  the weight  of a certain partial wave of the \ry electron is
\begin{equation}
{\cal C}_{n,l}(R)=\sum_{N,J}|C_{n,l,m}^J(R)|^2\qquad \text{satisfying} \qquad
\sum_{n,l}{\cal C}_{n,l}(R)=1. 
\label{eq:weight_l}
\end{equation}
For the sums of these  two previous expressions, it holds  $|l-N|\le J\le l+N$.

The  TURMs exist, if an APC can accommodate vibrational bound states. The corresponding 
vibrational spectrum is obtained 
by solving the Schr\"odinger equation associated to a given APC. We estimate the probability to create 
these TURMs by photoassociation via the Franck-Condon factor 
weighted with the $l=0,N=2$ admixture of the \ry wave function, 
\ie, $C_{n,l=0,N=2}^J (R)$, 
given by 
\begin{equation}
F_{v}=\int \left(\sum_{J,n} \Omega_{ns} C_{n,l=0,N=2}^J (R)\right) \chi^*_v(R) \psi_{scat}(R) R^2dR
\label{eq:FCF}
\end{equation}
with $\psi_{scat}(R)$ being the scattering wave function of the initial open channel and $\chi_v(R)$ the
vibrational wave function of the final APC, and
$ \Omega_{ns} $ the Rabi frequency for a Rydberg excitation to the $ns$ state. 
By assuming that the process is dominated by a single $ns$ \ry 
state, \ie, there is no significant admixture of different principle quantum numbers, then we can drop the sum over $n$.
In addition, we assume a constant transition dipole moment and a constant scattering wave function $\psi_{scat}(R)$. 
Also, to use the standard two-photon scheme for the \ry excitation, the final state should have a 
significant contribution of the $s$ or $d$-partial waves. This is estimated by the weight of a given partial 
wave of the \ry electron wave 
function  Eq.~\eqref{eq:weight_l} integrated over the vibrational wave function,  defined as 
\begin{equation}
W_{n,l}=\int \chi^*_v(R){\cal C}_{n,l}(R) \chi_v(R) R^2dR
\label{eq:weight_integrated_l}
\end{equation}
with $\sum_{n,l}{W}_{n,l}=1$. These integrated weights of the $s$ and $d$-partial waves should be 
approximately larger than $0.05$ in order to be able to create these molecules experimentally.  
Since  KRb is experimentally prepared either in its ground state or
in the $N=2$ excited state, we have also to analyze  the weight of a given rotational partial 
wave of the KRb wave 
function  Eq.~\eqref{eq:weight_N} integrated over the vibrational wave function,  defined as 
\begin{equation}
W_{N}=\int \chi^*_v(R){\cal C}_{N}(R) \chi_v(R) R^2dR
\label{eq:weight_integrated_N}
\end{equation}
with $\sum_{N}{W}_{N}=1$.
Finally, we also estimate the orientation, alignment and electric
dipole moment of the bound vibrational states by 
\begin{equation}
\langle\chi_v|X| \chi_v\rangle=\int \chi^*_v(R)X_{ryd}(R) \chi_v(R) R^2dR\qquad 
\label{eq:expectation_values_integrated_l}
\end{equation}
with $X=O,A,D$ for the orientation alignment and electric dipole moment, respectively.
The expressions of the $R$-dependent orientation alignment and electric dipole moment,
$O_{ryd}(R)$, $A_{ryd}(R)$, $D_{ryd}(R)$,  are
given in  Eq.~\eqref{eq:expec_coseno},  Eq.~\eqref{eq:expec_coseno2} and  Eq.~\eqref{eq:expec_dipole_moment}, respectively.

\section{Potential curves, Franck-Condon factors and Discussions}

\label{sec:results}
\begin{figure*}[t]
\begin{center}
    \includegraphics[scale=0.7]{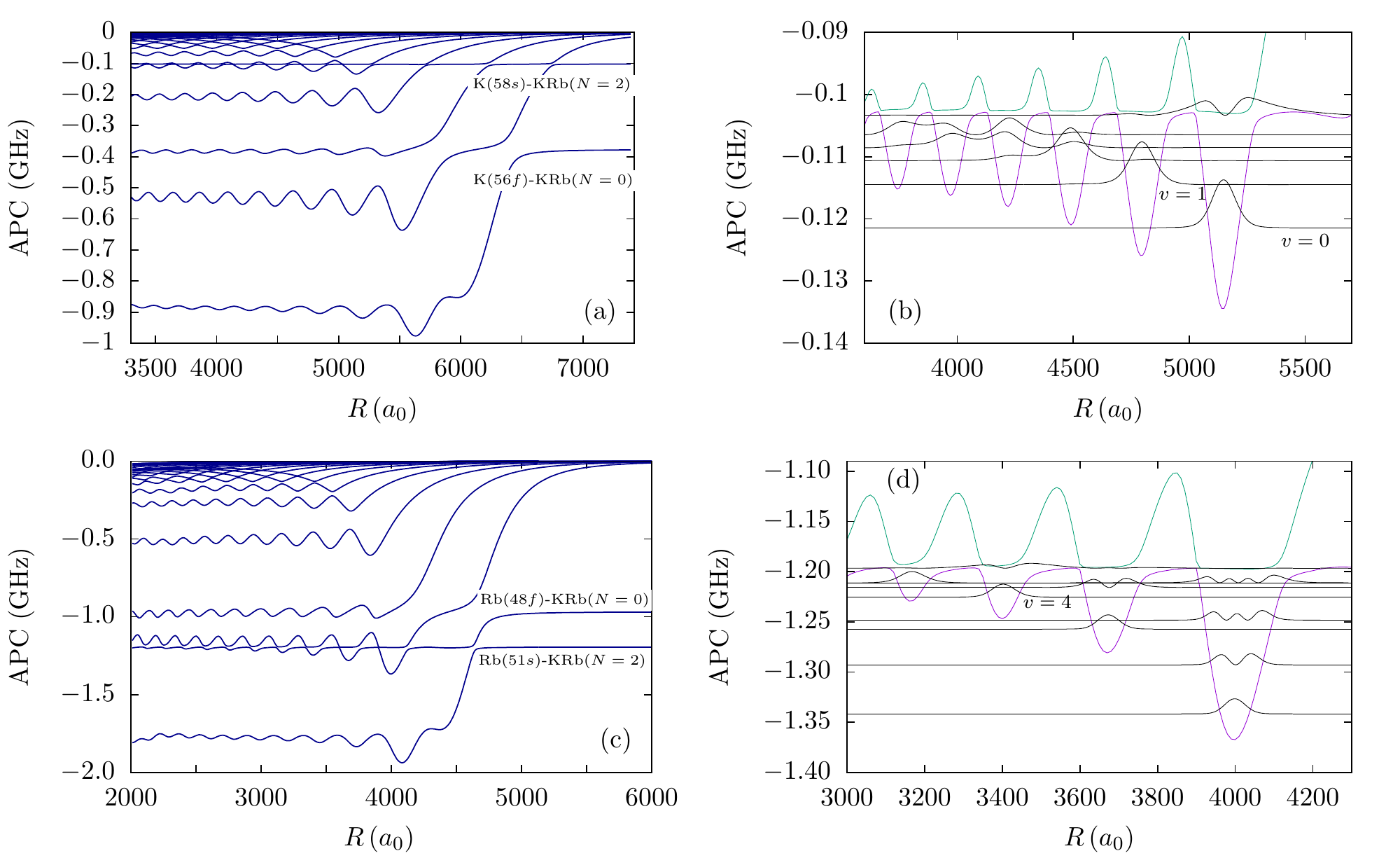} 
\caption{(Top panels) (a)  the $M_J=0$ APC curves for the K-KRb TURM,
evolving from K$(n=56,l\ge3)$- KRb(N=0) limit
and K$(58s)$-KRb(N=2) limit, 
(b) details of the APC curves,  near the K$(58s)$-KRb(N=2) dissociation threshold; (bottom panels) (c) the $M_J=0$ APC curves for the Rb-KRb TURM, evolving from Rb$(n=54,l\ge3)$- KRb(N=0) limit and Rb$(n=51s)$- KRb(N=2) limit, (d) details of the APC curves, near the Rb$(n=51s)$- KRb(N=2) dissociation threshold. The squared vibrational wave 
functions, shifted to the corresponding vibrational energies, are shown, for visualization and to illustrate the resolution in the vibrational spacing. Due to appreciable mixing of the K$(58s)$ and Rb($51s)$ with the respective degenerate manifolds, two-photon excitation to these levels can form the TURM. 
\label{fig:BOP_56}}
\end{center}
\end{figure*}
In this section, we describe the adiabatic electronic potential 
energy curves evolving from an $ns$ or $nd$ \ry state, which 
could be used experimentally in the protocol described above
to create a TURM. 
\subsection{K Rydberg excitation}

Let us start analyzing  the K-KRb   Rydberg TURM. 
The APCs of the TURM formed by K$(n=56,l\ge3)$ and the KRb in the ground state are presented 
in~\autoref{fig:BOP_56}~(a).  
Due to the oscillatory behavior of the Rydberg electron wave function, these APCs oscillate as the 
distance between the diatomic molecules and Rb$^+$ increases. The two lowest-lying APC
show potential wells with depths up to $1$~GHz. 
The K$(58s)$-KRb($N=2$) APC can be identified among these electronic  states
as the horizontal line with energy close to $-0.1$~GHz in~\autoref{fig:BOP_56}~(a).
In~\autoref{fig:BOP_56}~(b), we observe that this APC interacts most strongly with other electronic states evolving from K$(n=56,l\ge3)$-KRb($N=0$), near the avoided crossings. 
For this APC, the outermost well is the deepest one with a depth of approximately $\sim 30$~MHz. 
Most important, each of  these potential minima can accommodate at least one vibrational bound state, 
whose  wave functions  squared, shifted to the corresponding vibrational energies,  are shown
in~\autoref{fig:BOP_56}~(b).
As an example, we provide 
the energy spacing of the two lowest-lying vibrational levels, $\omega_{0-1}\sim 6.9$ MHz, see~\autoref{fig:BOP_56}~(b).
Within this APC, the polar diatomic molecule is neither oriented nor aligned, due to the small
contribution of odd rotational states to the total electronic 
wave function, see~\autoref{tab:vib_n56}.
At these vibrational levels, TURM possesses large permanent electric dipole moments.
This electric dipole moment is of the same order of magnitude as the one experimentally
observed for the ultralong range Rydberg molecules where the perturber is a ground state atom~\cite{Li1110,booth15} and here comes about from the near resonance of the Rydberg electron and KRb rotational energies.
\begin{table*}[h]
\begin{center}
\begin{tabular}{|c|c|c|c|c|c|}
\hline
$v$  & $E$ (GHz)& 
$\Delta E_{v,v+1}$ (MHz)&
$\langle\chi_v|O| \chi_v\rangle$& $\langle\chi_v|A| \chi_v\rangle$ & $\langle\chi_v|D| \chi_v\rangle$ \\
\hline
 0   &  $-0.12140$  &$6.91$&      $0.040$   &  $0.351$ &     $3656.07$\\
 \hline
  1  &  $-0.11449$&$3.85$ &       $0.039 $    &$0.362$ &    $3234.38$ \\
\hline
\end{tabular}
\caption{
For the  K$(58s)$-KRb($N=2$) APC: vibrational energies, vibrational spacing,  orientation, alignment, and electric dipole moment (in atomic units) of the two lowest lying vibrational states of the K$(58s)$-KRb($N=2$) APC in~\autoref{fig:BOP_56}~(b).}
\label{tab:vib_n56}
\end{center}
\end{table*}%

\begin{figure*}[t]
\begin{center}
 \includegraphics[scale=0.55,angle=0]{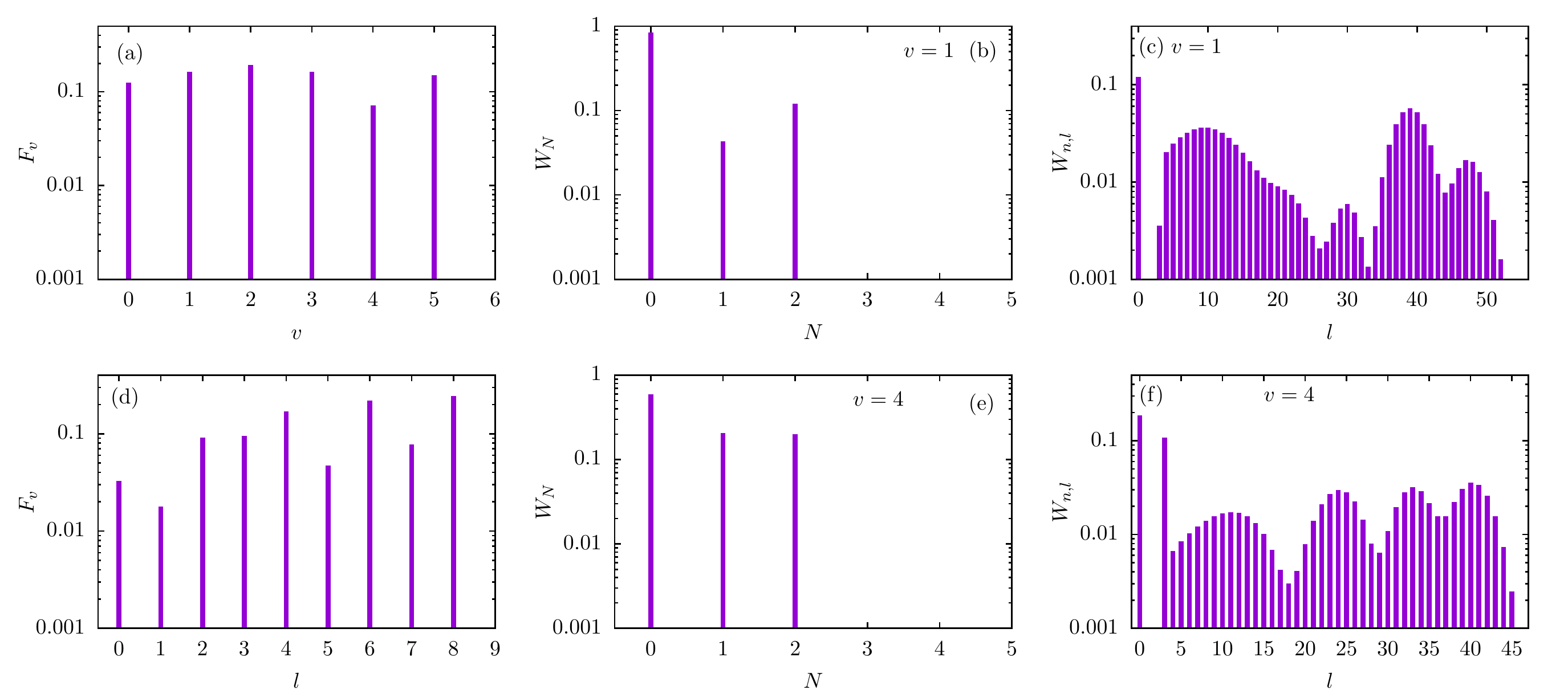} 
\caption{(Top panels) (a) Weighted Franck-Condon factors, Eq.~\eqref{eq:FCF} in the  electronic state evolving from K$(58s)$-KRb($N=2$);
(b) integrated weight of the rotational states Eq.~\eqref{eq:weight_N} of K-KRb TURM in the $\nu=1$ in the most outerwell in Fig.~\ref{fig:BOP_56}~(b); 
 (c) integrated  weight of the $l$-partial waves, Eq.~\eqref{eq:weight_l}, of K-KRb TURM in the $\nu=1$ in the most outerwell in Fig.~\ref{fig:BOP_56}~(b); (bottom panels) 
 (d) F$_\nu$ for vibrational states in the APC evolving from Rb$(51s)$-KRb($N=2$);
(e) W$_N$ for rotational levels of the Rb-KRb TURM in the $\nu=4$ level in Fig.~\ref{fig:BOP_56}~(d); 
 (f) W$_{nl}$ of Rb-KRb TURM in the $\nu=4$ in Fig.~\ref{fig:BOP_56}~(d). 
 \label{fig:integrated_weigth_l}}
\end{center}
\end{figure*}

\autoref{fig:integrated_weigth_l}~(a) shows the
weighted Franck-Condon factors~\eqref{eq:FCF} for the six vibrational bound  states of the APC  K$(58s)$-KRb($N=2$) in~\autoref{fig:BOP_56}~(b). 
All the bound vibrational states present similar values for this
weighted Franck-Condon factor, and its dependence on the vibrational
quantum number illustrates their nodal structure.
The integrated weights of the rotational states of KRb for the vibrational state $v=1$, \ie, the vibrational state located on the second outermost well of the APC, 
are shown in~\autoref{fig:integrated_weigth_l}~(b).
Three rotational partial weights $N=0,1,$ and $2$ contribute to the total wave function, the contribution from the $N=0$ rotational state  being
the dominant one. 
The small contribution of the $N=1$ rotational state justifies the lack of orientation of KRb with the APC  K$(58s)$-KRb($N=2$). 
The  weights of the partial waves of the \ry electron wave function 
integrated with the vibrational wave functions, see  Eq.~\eqref{eq:weight_integrated_l}, for 
the vibrational state $v=1$ of the APC are presented in~\autoref{fig:BOP_56}~(c). 
For this vibrational bound state, many partial waves with $l\ge 3$ have an integrated weight larger 
than $1\%$,  indicating that they have a significant contribution from the K$(56, l>3)$-KRb($N=0$) APC. 
The avoided crossing between the K$(58s)$-KRb($N=2$) APC and the APC evolving from 
K$(56, l>3)$-KRb($N=0$) is manifested in the significant contribution of the $s$-partial wave ($l=0$), 
\eg, the integrated weight 
over the vibrational wave function for this $v=1$ vibrational  bound state is 
$12\%$.
The large weighted Franck-Condon factors, combined with  the large
contributions from the \ry state K$(58s)$ guarantee that these molecules 
can be created by the standard two-photon absorption scheme if KRb is in its rotational states $N=0$ or $N=2$. 

\subsection{Rb Rydberg excitation}
\begin{figure*}[t]
\begin{center}
     \includegraphics[scale=0.85]{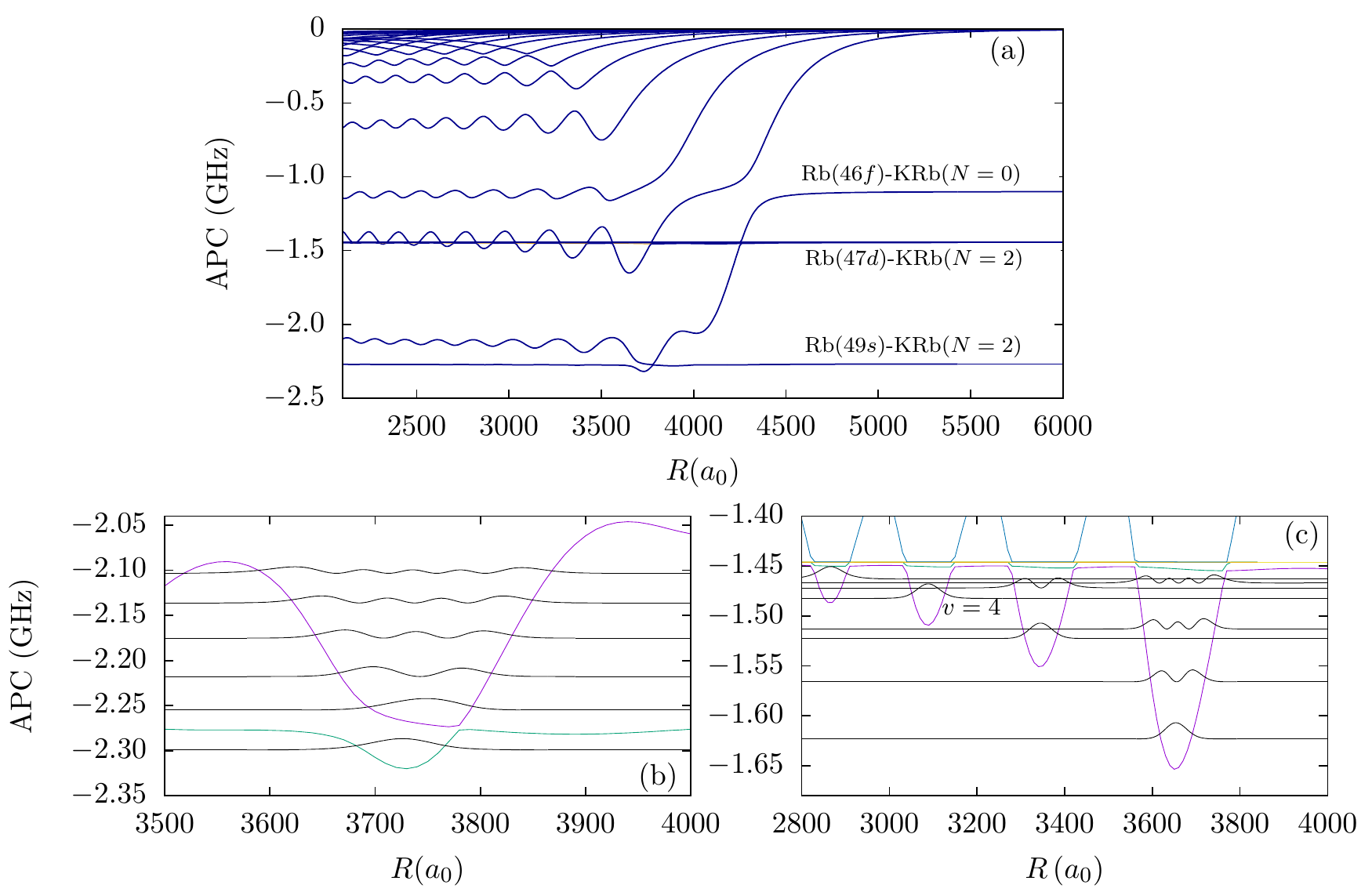} 
\end{center}
\caption{For the Rb-KRb   \ry  TURM,  (a) adiabatic electronic potential curves  for  $M_J=0$
evolving from the \ry state R$(n=46,l\ge3)$ and the KRb in the ground state, Rb$(47d)$-KRb(N=4), 
and  Rb$(49s)$-KRb($N=2$).
(b) Details of the APC evolving from  Rb$(49s)$-KRb($N=2$)  and the neighboring level evolving from 
Rb$(n=46,l\ge3)$-KRb($N=0$), together with the square of the vibrational wave functions shifted to the 
corresponding  vibrational energies.
(c)  Details of the APC evolving from  Rb$(47d)$-KRb($N=4$)  and the neighboring level evolving from 
Rb$(n=46,l\ge3)$-KRb($N=0$). The square of the vibrational wave functions shifted to the corresponding vibrational energies are also shown. 
Note that the square of the vibrational wave functions  is
not plotted up to scale: they have been magnified for better visibility.
\label{fig:BOP_Rb_46}}
\end{figure*}

We explore now the electronic structure of the TURM formed by exciting the rubidium atom, \ie, the Rb-KRb  TURM.
The APCs evolving from the \ry degenerate manifolds Rb$(n=46,l\ge3)$  and  neighbouring 
quantum-defect states are presented in~\autoref{fig:BOP_Rb_46}.
Among these APCs, we identify electronic states satisfying the conditions indicated above, which 
are those evolving from Rb$(47d)$-KRb($N=4$)  and Rb$(49s)$-KRb($N=2$), see the horizontal lines, on 
the scale of the figure, with energies $-1.45$~GHz and  $-2.28$~GHz, respectively. 

\begin{figure*}[t]
\begin{center}
 \includegraphics[scale=0.55,angle=0]{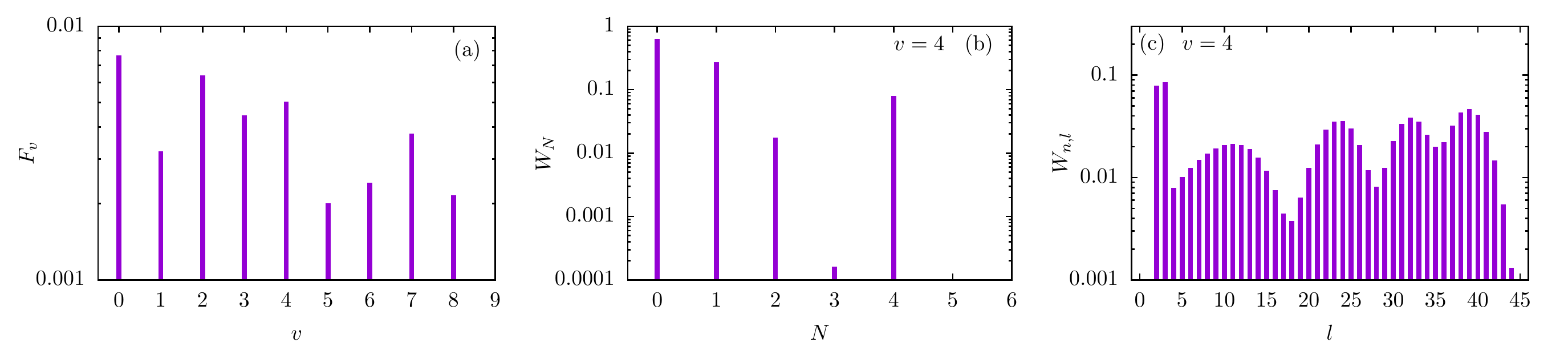} 
\caption{For the R-KRb \ry TURM, (a) weighted Franck-Condon factors Eq.~\eqref{eq:FCF} for the vibrational bound states of the electronic state evolving from Rb$(48d)$-KRb($N=4$);
 (b) integrated weight of the rotational states of KRb on
 the electronic wave function Eq.~\eqref{eq:weight_integrated_N} of the
 electronic wave function  Eq.~\eqref{eq:weight_integrated_N} for
 the vibrational state  $v=1$; 
 (c) integrated  weight of the \ry $l$-partial waves in  the electronic 
 wave function  Eq.~\eqref{eq:weight_integrated_l} for the vibrational state $v=1$.\label{fig:integrated_weigth_l_RKRb_48_35}}
\end{center}
\end{figure*}

The avoided crossing with the Rb$(47d)$-KRb($N=4$) APCs, gives rise to deep potential wells
having several vibrational bound states, whose squared wave functions are presented in~\autoref{fig:BOP_Rb_46}~(b). 
The weighted Franck-Condon factors of the nine vibrational bound states of this potential are shown in~\autoref{fig:integrated_weigth_l_RKRb_48_35}~(a).
For all these states, the weighted Franck-Condon factors are small, and 
again their dependence on the vibrational quantum number resembles the
nodal structure of the vibrational wave function.
For the $v=4$ vibrational state, we present the integrated weights of the
rotational wave function in~\autoref{fig:integrated_weigth_l_RKRb_48_35}~(b). There are four
field-free rotational states of KRb contributing to the wave function,  the $N=0$ contribution  being the dominant one due to the avoided crossing. Within this vibrational bound state  KRb is anti-oriented.
The integrated weights of the partial waves of the \ry electron wave function show a dominant 
contribution for the $l\ge 3$ partial waves for the low-lying vibrational states. This indicates that 
these vibrational bound states are mainly from the APC evolving from the degenerate manifold Rb($n=46,l\ge3$) because they are energetically far from the avoided crossing. 

For higher excited vibrational states, \ie, as their energies  approach the avoided crossing region,
the contribution of the $d$-wave increases and becomes the largest one. 
For instance, the integrated weights for the $v=4$ vibrational level are presented 
in~\autoref{fig:integrated_weigth_l_RKRb_48_35}~(c). The wave function of this  fifth vibrational level has 
around $7.9\%$  contribution of the $d$-wave, which becomes even larger for higher excitations.
This bound state has a vibrational spacing of $10$~MHz with respect to the upper neighbouring vibrational 
states, see ~\autoref{tab:vib_Rb_n48}, and of $30$~MHz with respect to the lower lying one.
In addition, it has a large
electric dipole moment, but it is weakly anti-oriented and aligned.
We can conclude that the TURM molecule could be experimentally created from the quantum-defect 
state Rb$(47d)$ if KRb is in the excited rotational state KRb($N=4$). 

\begin{figure*}[t]
\begin{center}
 \includegraphics[scale=0.6]{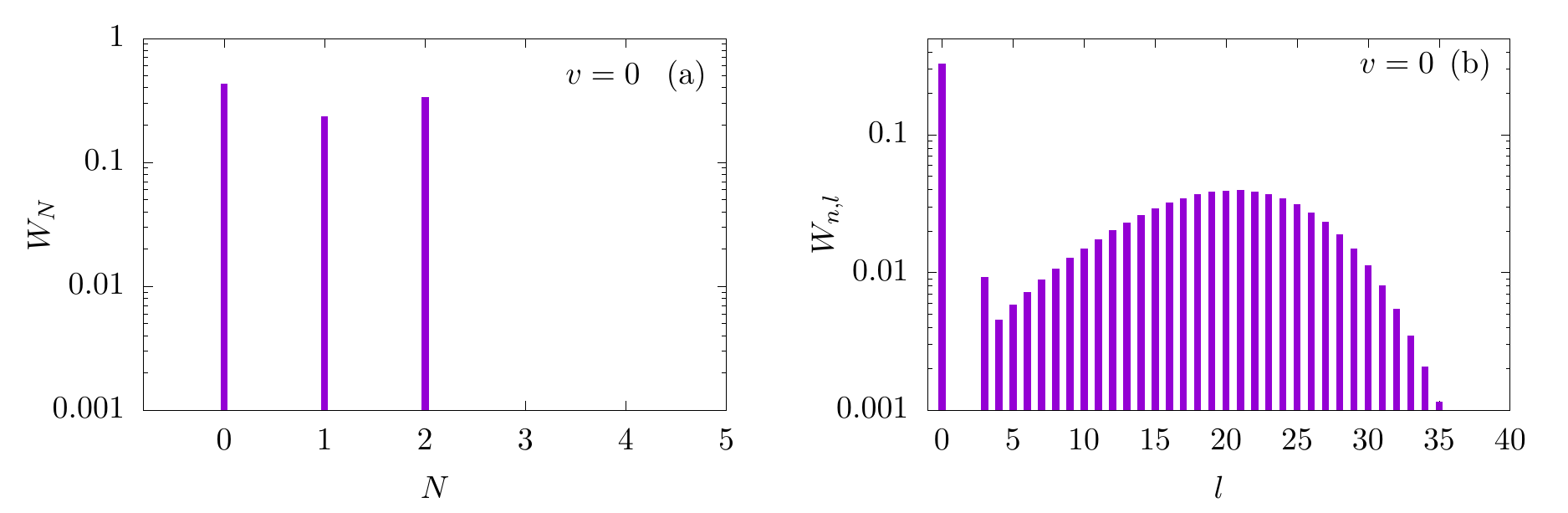}
\caption{For the $v=0$ vibrationl state in the electronic state evolving from Rb$(49s)$-KRb($N=2$) \ry TURM, 
(a) integrated weight of the rotational states of KRb and 
(b) integrated weight of the $l$-partial waves in  the  electronic wave function
of this  vibrational state.  \label{fig:weights_Rb_KRb}
}
\end{center}
\end{figure*}

The two APCs evolving from Rb$(49s)$-KRb($N=2$) and the neighbouring state from the  
Rb$(n=46,l\ge3)$-KRb($N=0$) manifold, see~\autoref{fig:BOP_Rb_46}~(c), can accommodate up to
six vibrational bound states with energy spacings ranging from $50$~MHz to $35$~MHz. The broad 
avoided crossing between these two APCs gives rise to a non-adiabatic coupling, and, as a 
consequence, only the lowest-lying vibrational state is stable, whereas the excited ones, \ie, those lying
on the upper APC of~\autoref{fig:BOP_Rb_46}~(c), will decay and
the TURM will dissociate. 
The weighted Franck-Condon factor to this lowest-lying vibrational state is $0.4$, implying a favorable photoassociation rate.
For this vibrational state, the integrated weight of the rotational
state of KRb is 
presented in~\autoref{fig:weights_Rb_KRb}~(a). This vibrational state has a significant contribution of the field-free
$N=0$ and $N=2$ rotational states, but the KRb shows a moderate alignment,  see~\autoref{tab:vib_Rb_n48}. Due to the contribution of the 
$N=1$ rotational state, the diatomic 
molecule is weakly oriented, see also~\autoref{tab:vib_Rb_n48}.  
The integrated weights of the partial waves of the \ry electron wave function are presented 
in~\autoref{fig:weights_Rb_KRb}~(b), where the $s$-wave has the dominant 
contribution of $31\%$. 
Due to these properties, we can conclude that the TURM molecule  could 
be experimentally created from the quantum-defect state Rb$(49s)$ if  
KRb is in  the $N=2$ rotational state.

For completeness, we present in~\autoref{fig:BOP_56}~(c) the electronic structure of the TURM formed from  the \ry degenerate manifolds Rb$(n=48,l\ge3)$  and the ground-state diatomic molecule. 
In this configuration, 
we encounter the adiabatic electronic state evolving from the  Rb$(51s)$-KRb($N=2$), which suffers 
an avoided crossing with an APC evolving from  Rb$(n=48,l\ge3)$-KRb($N=0$), 
see~\autoref{fig:BOP_56}~(d). 
The potential wells accommodate $8$ vibrational bound states, whose wave functions
are shown~\autoref{fig:BOP_56}~(d). As in the previous cases, 
we present in~\autoref{fig:integrated_weigth_l}~(d) the weighted Franck-Condon factor of these vibrational states. 
The low-lying vibrational bound levels 
are far from the avoided crossing,  and they do not show a significant contribution of the $s$-wave \ry 
wave function. As the energy of these vibrational states increases, the avoided crossing 
strongly affects their properties. In particular, for the fourth vibrational state,
the integrated weight of the \ry partial wave to the electronic wave function are presented in~\autoref{fig:integrated_weigth_l}~(f).
The \ry $s$-wave has the dominant contribution with $19\%$, which is even larger for higher excited 
vibrational states. In addition, the integrated weights of the KRb 
rotational states to the $v=4$ vibrational bound state are shown in~\autoref{fig:integrated_weigth_l}~(e). Within this vibrational state  
the KRb is antioriented. 
The properties of this vibrational bound state are shown in~\autoref{tab:vib_Rb_n48}.

\begin{table*}[h]
\begin{center}
\begin{tabular}{|c|c|c|c|c|c|c|c|}
\hline
APC &$v$  & $E$ &  $\Delta E_{v,v+1}$&$\langle\chi_v|O| \chi_v\rangle$& $\langle\chi_v|A| \chi_v\rangle$ & 
$\langle\chi_v|D| \chi_v\rangle$\\
& & [GHz] &[MHz] & & & \\
\hline
Rb$(47d)$-KRb($N=4$)  & 4   & $-1.4830$   &$10.7$&    $ -0.1122$   &     $0.4347$     &      $2006.2$ \\

\hline
Rb$(49s)$-KRb($N=2$)  & 0   & $-2.2992$   &$21.7$&    $0.2284$   &     $0.4173$     &      $1857.1$\\

\hline

 Rb$(51s)$-KRb($N=2$) & $4$    &  $-1.2255$ & $9.57$  &   $-0.0973$    &  $0.4348$    & $1877.9$   \\
 \hline

\end{tabular}
\caption{For the Rb-KRb \ry TURM:  vibrational energies, vibrational spacing, 
orientation, alignment, and electric dipole moment (in atomic units) for selected vibrational bound states of different APCs.}
\label{tab:vib_Rb_n48}
\end{center}
\end{table*}

\section{Conclusions}
\label{sec:con} 
We have investigated  triatomic ultralong-range Rydberg 
molecules formed by a \ry  rubidium or potassium atom and a KRb diatomic rotational molecule. We have explored a regime where the dipole
interaction of the \ry electron with the diatomic polar molecule induces a coupling 
between the quantum defect \ry states and the nearest degenerate hydrogenic manifold. 
Due to this induced coupling,  the adiabatic potential wells evolving from the
hydrogenic like \ry manifold acquire $s$- and $d$-wave admixture from the corresponding 
\ry quantum defect states.
As a consequence, 
an experimental protocol with the conventional two-photon excitation can be employed to create TURM  \cite{bendkowsky09,Li1110,booth15}
could be used to prepare the \ry triatomic Rydberg molecules 
described here. To get a measurable photo-association rate, the densities of the atomic and molecular gases must be large enough such that there is a reasonable probability of finding on average one molecule inside the Rydberg orbit.  For Rb-KRb TURM  photo-association, a relatively dilute gas of Rb with density $\sim 1 \times 10^{12}$ cm$^{-3}$ overlapping with a gas of KRb molecules in their electronic and vibrational ground state prepared in the $N=2$ rotational state with a density of $2\times 10^{12}$ cm$^{-3}$ \cite{DeMarco2019} has on average $\sim 0.1$ KRb molecules within a Rb Rydberg orbit.
Since it is based on the dipole-induced coupling between different \ry states, this experimental protocol is ubiquitous and could be applied
to any pair of species forming these ultralong-range \ry molecules. 
Optical tweezers may provide another platform to create the TURM. Overlap of single molecule trap \cite{Anderegg1156} and atom tweezer \cite{Bernien,wilson2019trapped}, which can be excited into Rydberg states, offer the opportunity to form triatomic Rydberg molecules at the right separation.

\section*{Acknowlegments}

R.G.F. gratefully acknowledges  financial support by the Spanish Project No. FIS2017-89349-P 
(MINECO), and by the Andalusian research group FQM-207. This study has been partially 
financed by the Consejer\'{\i}a de Conocimiento, Investigaci\'on y Universidad, Junta de 
Andaluc\'{\i}a and European Regional Development Fund (ERDF), Ref. SOMM17/6105/UGR. 
P.S. acknowledges financial support in the framework of PIER Hamburg-MIT/Seed 
Projects provided by the Hamburg Ministry of Science, Research and Equality (BWFE).


\bibliographystyle{apsrev4-1}  
\bibliography{TriMol}

\end{document}